\documentclass{article}
\begin{document}

\title{ Relativistic Brownian Motion in 3+1 Dimensions}

\author{O. Oron and L.P. Horwitz \thanks{Also at Department of Physics, Bar Ilan University, Ramat Gan 529000, Israel}
\\
School of Physics and Astronomy \\
Raymond and Beverly Sackler Faculty of Exact Sciences \\
Tel Aviv University, Ramat Aviv 69978, Israel}
 \date{15 September, 2003}
\maketitle
\begin{flushright}TAUP 2752-03
\end{flushright}
 
\begin{abstract} 
 We solve the problem of formulating Brownian motion in a
relativistically covariant framework in $3+1$ dimensions.
We obtain covariant Fokker-Planck equations with (for the isotropic
case) a differential operator of invariant d'Alembert form. Treating the
spacelike and timelike fluctuations separately in order to maintain the covariance property, we show that it is
essential to take into account the analytic continuation of
``unphysical'' fluctuations. 
We discuss the notion of locality in this framework and possible implications for entangled states. 
\end{abstract}        
\section{ Introduction}
The ideas of stochastic processes originated in the second half of the
 19th century in thermodynamics, through the manifestation of the
 kinetic theory of gases. In 1905 A. Einstein \cite{Einstein} in his paper on Brownian
 motion provided a decisive breakthrough in the understanding of the
 phenomena. Moreover, it was a proof convincing physicists of the
 reality of atoms and molecules,  the motivation for Einstein's
 work. It is interesting to note that Einstein predicted the so
 called Brownian motion of suspended microscopic particles not knowing
 that R. Brown first discovered it in 1827 \cite{Brown}. 
 The resemblance of the Schr\"{o}dinger equation to the diffusion equation had lead physicists (including Einstein and Schr\"{o}dinger) to attempt to connect quantum mechanics with an underlying stochastic process, the Brownian process.     
\par Nelson \cite{Nelson1966}, in 1966, constructed the Schr\"odinger equation from
an analysis of Brownian motion by identifying the forward and backward
average velocities of a Brownian particle with the real and imaginary
parts of a wave function. He pointed out that the basic process
involved is defined non-relativistically, and can be used if
relativistic effects can be ``safely'' neglected. The development
of a relativistically covariant formulation of Brownian motion could
 therefore
provide some insight into the structure of a relativistic quantum theory.
 \par Nelson pointed out that the formulation of his stochastic
mechanics in the context of general relativity is an important open
 question \cite{Nelson1985}.  The Riemannian metric
spaces one can achieve, in principle, which arise due to nontrivial
correlations between fluctuations in spatial directions, could,  in the
 framework of a covariant theory of Brownian motion, lead to
spacetime pseudo-Riemannian metrics in the structure of diffusion and
Schr\"odinger equations.
Morato and Viola \cite{65} have recently constructed a relativistic
 quantum equation for a free scalar field. They assumed the
 existence of a 3D (spatial) diffusion in a comoving frame, a
 non-inertial frame in which the average velocity field of the
 Brownian particle (current velocity) is zero. In this frame the
 location of the Brownian particle in space  experiences Brownian
 fluctuations parametrized by the proper time of the comoving
 observer. They interpreted possible negative
 0-component current velocities with what they called `rare events',
 which are time reversed Brownian processes (a peculiarity arising in
 the relativistic treatment). The equation they achieved this way is
 approximately the Klein-Gordon equation. It is important to note that
 in the inertial frame they do not obtain a normal diffusion. This is due to the
 fact that their process is stochastic only
 in three degrees of freedom and therefore is not covariant. In this paper we shall study a manifestly covariant form of Brownian motion.
 \par Nelson himself \cite{Nelson1985} argued that Markov processes may lead to inconsistencies (which he demonstrated in entangled systems such as E.P.R \cite{E.P.R})unless a non-local theory is allowed, therefore according to his own     preference for a local theory he suggested to consider non-Markovian theories.  We shall treat this problem in our concluding section.       
\par In a previous work {\cite{valeri}} we introduced a new approach to the formulation of relativistic Brownian motion in 1+1 dimensions. The process we formulate is a straightforward generalization of the standard one dimensional diffusion to 1+1 dimensions (where the
 actual random process is thought of as a `diffusion' in the time direction as well as in space), in an inertial frame. The equation achieved is
 an exact Klein-Gordon equation. It is a relativistic generalization
 of Nelson's Brownian process, the Newtonian diffusion. In this paper we review the relativistic Brownian process in 1+1 dimensions {\cite{valeri}} where the inclusion of both spacelike and timelike motion for the Brownian particle (event) is considered; if the timelike motion is considered as ``physical'' the ``unphysical'' spacelike motion is represented (through analytic continuation) by imaginary quantities. We extend the treatment to 3+1 dimensions using appropriate weights for the imaginary representations. The extension of the process to a \textit{general} covariant form will be carried out in a succeeding paper. The complete formalism then can be used to construct relativistic general covariant diffusion and Schr\"odinger equations with pseudo-Riemannian metrics which follow from the existence of nontrivial correlations between the coordinate random variables.
\par Finally, we discuss the possible implications of the process we consider (i.e. a relativistic stochastic process with Markov property which preserves macroscopic Lorentz covariance) on the entangled system, where we claim that though fluctuations which exceed the velocity of light occur the macroscopic behavior dictated by the resulting Fokker-Planck equation is local.

\section{The Problem of Assigning a Dynamical Evolution Parameter for the Relativistic Brownian Process and the Stueckelberg Formalism}
\bigskip
\bigskip
\par Brownian motion, thought of as a series of ``jumps'' of a
particle along its path, necessarily involves an ordered sequence.  In
the nonrelativistic theory, this ordering is naturally provided by the
Newtonian time parameter.  In a relativistic framework, the Einstein
time $t$ does not provide a suitable parameter. If we contemplate
jumps in spacetime, to accommodate a covariant formulation, a possible
spacelike interval between two jumps may appear in two orderings in
different Lorentz frames. The introduction of proper time as a
parameter for the RBP (Relativistic Brownian Process) is not adequate
since in this
 case the second order correlations in the simplest case (i.e. for an isotropic homogeneous process with a diffusion constant $\sigma^2$ ) have the form:
\begin{equation} 
E(\Delta x^\mu \Delta x^\mu)=2 \sigma^2 \Delta s \label{6.1.1}
\end{equation}
for each $\mu$; however, summing over $\mu$,
\begin{equation} E(\Delta x_\mu \Delta x^\mu)\equiv \Delta s^2 \propto \Delta s,
 \label{6.1.2}\end{equation}
where the first equality is by the definition of proper time and the
 second equality is due to the Brownian property expressed in
 Eq.~(\ref{6.1.1});
 there is an obvious contradiction. We therefore adopt the invariant
parameter $\tau$ as the dynamical variable for the Brownian process, first suggested in 1941 by E.C.G Stueckelberg \cite{Stueckelberg}. 
 The introduction of the notion of an invariant time $\tau$ permitted the discussion of world lines not monotonic in the ordinary (Einstein) time, $t$. On such world
 lines two points may have the same $t$ coordinate and represent the
 occurrence of two particles existing at the same time. The time coordinate, $t$, may increase or decrease as $\tau$ evolves (just as a particle in Newtonian mechanics may move in the positive or negative spatial directions with time) representing particles and antiparticles respectively. In this way Stueckelberg was able to describe pair-creation and annihilation on a classical level and write a relativistic Schr\"odinger equation (with four coordinates of spacetime and the invariant time as a parameter).       
\par In 1973 Horwitz and Piron \cite{33} developed this concept, suggesting that
 all
 physical systems evolve through one universal invariant time, not affected by interaction or dynamical coordinate transformations, just as in the Newtonian theory. The theory formally resembles Newton's theory with the most significant difference that Euclidian space is replaced by Minkowski spacetime.
\par The point particle, in Newtonian mechanics, represented by its
 spatial coordinates at each given time, is replaced by a new concept,
 the {\it event}; it is an object, evolving in spacetime, represented by four coordinates at each given (universal) time
 $\tau$. The collection of the event's (four) positions in spacetime
 at all (universal) times forms the worldline. The time-coordinate of
 an event in a given inertial frame is the time in which, according to
 the clocks of this frame the event is detected, though it could have
 occurred, in principle, at a different (universal) time, $\tau$. In
 particular let us consider a parabolic worldline (figure 1), describing pair annihilation, at a given time $t_{OB}$, such that the line $t=t_1$ intersects the worldline twice. At the time $t_1$  two events will be detected on the frame's clocks, one corresponding to a particle and the other to an antiparticle, each occurring at  different (universal) time $\tau$. The interpretation of the event going backwards in time as the antiparticle was given first by Stueckelberg and was later used by Feynman; it is now an accepted concept. However, in the parabolic worldline there are segments in which the event goes faster than light (either forward or backward in $t$). Do these segments have any physical representation? Should they be included in a physical theory? As we shall see in this paper the answer for the second question is positive in the formulation of our relativistic Brownian process, and the occurrence of this state of motion is dictated by the demand of achieving Lorentz invariance (more explicitly the d'Alembert) operator in the relativistic diffusion equation.     
\begin{center}
\end{center}
{\bf figure 1 : Particles and Events.} A particle/antiparticle corresponds to a world line segment formed
of a trajectory of an event whose time coordinate is monotonically
increasing/decreasing with $\tau$. At $t_1$ a particle and an
anti-particle may be detected, both generated by an event $B$. The particle/antiparticle corresponds  to
the (left/right) branch of the parabola (within the forward/backward light cones) where the time coordinate is
increasing/decreasing with $\tau$. The segment of motion outside of the lightcones are tachyonic (spacelike) and are required by continuity of the worldline. These two points occur at different
$\tau$ but at the same $t$. At times
$t>t_{OB}$ (the turning point), the event $B$ does not exist therefore the
worldline generated by event $B$ corresponds to a pair annihilation
occurring at $t=t_{OB}$. An event, say, $C$ may generate a worldline corresponding to
a pair creation occurring at $t=t_{OC}$ and another event $A$ may generate a monotonic
worldline therefore corresponding to a single particle.

\section{The Negative Correlation Problem }                                             
\par A second fundamental difficulty in formulating a covariant theory
of Brownian motion lies in the form of the correlation function of the random variables of spacetime. The correlation function for the isotropic (non-relativistic) Wiener \cite{Wiener} process, is given by, 
\begin{eqnarray}
<\Delta x^i \Delta x^j>=\sigma^{ij}dt \nonumber \\
\sigma^{ij}=2 \sigma^2\delta^{ij} \nonumber \\
i,j=1,2,3 \label{wiener} 
\end{eqnarray}
where the $\delta^{ij}$ generates the Euclidian structure for the manifold on which the Brownian evolution take place \cite{Nelson1967}.  The straightforward covariant generalization to the relativistic case is,
\begin{equation} 
<dw_\mu(\tau)dw_\nu(\tau')>=
\begin{array}{cc}
 0   & \tau \not= \tau'\\
  2 \alpha^2 \eta_{\mu\nu}d\tau &  \tau=\tau'  \\
\end{array}
  \label{6.1.3}
\end{equation}    
where  $\eta^{\mu \nu}=\textrm{diag}(-1,1,1,1)$ is the Minkowski metric, and therefore \break $<dw_0(\tau)dw_0(\tau)>\,\, <0,$
which is impossible. Let us consider, however, a process which is physically restricted to only to spacelike or timelike jumps. One may argue that Brownian motion in spacetime should be a generalization of the non-relativistic problem,
constructed by observing the non-relativistic process from a moving
frame according to the transformation laws of special relativity.
Hence the process taking place in space in the
non-relativistic theory would be replaced by a spacetime process
in which the Brownian jumps are spacelike.  The pure time
(negative) self-correlation does therefore not occur. In order to
meet this requirement, we shall use a coordinatization in terms of
generalized polar coordinates which assures that all jumps are
spacelike. Consider for example a relativistic Brownian
probability density of the form $e^{-{\mu^2\over a
d\tau}}$, where $\mu$ is the invariant spacelike interval of the
jump. This is a straightforward generalization of the standard
Brownian process in 3D, which is generated by a probability
density of the form $e^{-r^2 \over a dt}$, where r is the rotation
invariant (i.e. the vector length)and $a$ is proportional to the
diffusion constant. We shall refer to this function as the
relativistic Gaussian.
 As we shall see, a Brownian
motion based on purely spacelike jumps does not, however, yield the
correct form for an invariant diffusion process.  We must therefore
consider the possibility as well that, in the framework of
relativistic dynamics, there are timelike jumps.  The corresponding
distribution would be expected to be of the form
$e^{-{\sigma^2\over b d\tau}}$,
where $\sigma$ is the invariant interval for the timelike jumps, and $b$ is
some constant.
 By suitably
weighting the occurrence of the spacelike process (which we take for
our main discussion to be ``physical'', since its nonrelativistic
limit coincides with
the usual Brownian motion) and an analytic continuation of the
timelike process, we show that
one indeed obtains a Lorentz invariant Fokker-Planck equation in
which the d'Alembert operator appears in place of the Laplace
operator of the 3D Fokker-Planck equation. One may, alternatively,
consider the timelike process as ``physical''(as might emerge from a microscopic model with scattering) and analytically
continue the spacelike (``unphysical'') process to achieve a
d'Alembert operator with opposite sign.
\bigskip
\bigskip
\section{  Brownian motion in 1+1 dimensions}
\bigskip
\bigskip
\par We consider a Brownian path in $1+1$ dimensions generated by a
 stochastic differential (analogue to the Langevin equation {\cite{Langevin}} and Smoluchowsky process \cite{Smoluchowsky}), of the form:
\begin{equation} dx^\mu (\tau) =  \beta^\mu(x(\tau))d\tau + dw^\mu (\tau),
 \label{6.2.1}\end{equation}
where $dw$ is a random process, which is a relativistic generalization
of the Wiener process, whose properties will be defined later, and
$\beta^\mu$
 is a deterministic field (the drift).
\par We start by considering the second order term in the series
expansion of a function of position of the particle on the world line,
$f(x^\mu(\tau) + \Delta x^\mu)$,
involving the operator
\begin{equation}{\cal O}={\Delta x}^{\mu}{\Delta x}^{\nu}{{\partial} \over
 {\partial}{x}^\mu}{\partial \over \partial {x}^\nu}. \label{6.2.2}\end{equation}
We have remarked that one of the difficulties in describing Brownian
motion in spacetime is the possible occurrence of a negative value for
the second moment of some component of the Lorentz four vector random
variable. If the Brownian jump is timelike, or spacelike, however, the
components of the four vector are not independent, but must satisfy
the timelike or spacelike constraint.  Such constraints can be
realized by using parameterizations for the jumps in which they
are restricted geometrically to be timelike or spacelike.
  We now separate the random jumps into space-like jumps and time-like
 jumps
accordingly, i.e., for the spacelike jumps,
\begin{equation} \Delta w^1=\pm \mu \cosh{\alpha} \,\, , \,\, \Delta w^0 =\mu
 \sinh{\alpha}\label{6.2.3}\end{equation}
and for the timelike jumps,
\begin{equation} \Delta w^1=\sigma \sinh{\alpha} \,\, , \,\, \Delta w^0 =\pm \sigma \cosh{\alpha} \label{6.2.3'}\end{equation}
Here we assume that the two sectors have the same distribution
on the hyperbolic variable. We furthermore assume that $\mu$,$\sigma$
are generated by a relativistic Gaussian distribution, working in a
Lorentz frame where the $\alpha$ distribution is assumed to be independent of $\mu,\sigma$ and is uniformly distributed on the
restricted interval  $[-L,L]$ (see discussion below) where $L$ is arbitrary
large. Therefore, in this frame $<\Delta w^\mu>$ is 0 (this is true in all frames; see discussion in Section 5) and we pick a
normalization such that (for any component) $<\Delta w^n> \propto \Delta \tau^{n\over 2}$
so to first order in $\Delta \tau$ the
 contribution to $<\cal O>$ comes only from $<\Delta w^\mu \Delta w^\nu>$.
\par  For a particle experiencing space-like
jumps only,  the operator ${\cal O}$ takes the following
 form:
\begin{equation} {\cal O}_{spacelike}= \mu^2[{\cosh^2}{\alpha}{\partial^2 \over \partial
x^2}+2\sinh\alpha \cosh\alpha {\partial^2 \over {\partial x \partial
t}}+
\sinh^2\alpha{\partial^2 \over \partial t^2}] \label{6.2.4}\end{equation}
If the particle undergoes time-like jumps only the operator
${\cal O}$ takes the form:
\begin{equation} {\cal O}_{timelike}= \sigma^2 [\sinh^2 \alpha{\partial^2 \over \partial
x^2}+2\sinh\alpha \cosh\alpha {\partial^2 \over {\partial x \partial
t}}+\cosh^2\alpha {\partial^2 \over \partial t^2}] \label{6.2.5}\end{equation}

\par  Since $\mu,\sigma$ and $\alpha$ are random processes, the average value of
 the operator ${\cal O}$ is the sum of the two averages of Eq.~(\ref{6.2.4}) and Eq.~(\ref{6.2.5}).
 A difference between these two averages, leading to the d'Alembertian operator can only be obtained by considering the analytic continuation of
the timelike process to the spacelike domain, choosing
$\mu^2=-\sigma^2$. 
\par This procedure is analogous to the effect,
well-known in relativistic quantum scattering theory, of a
physical process in the crossed ($t$)channel on the observed
process in the direct ($s$) channel.  For example, in the LSZ
formulation of relativistic scattering in quantum field theory [e.g. \cite{86}], a
creation operator in the ``in'' state may be moved to the left in the
vacuum expectation value expression for the $S$-matrix, and an
annihilation operator for the ``out'' state may be moved to the
right.  The resulting amplitude, identical to the original one in
value, represents a process that is unphysical; its total
``energy'' (the sum of four-momenta squared) now has the wrong
sign.  Assuming that the $S$-matrix is an analytic function, one
may then analytically continue the energy-momentum variables to
obtain the correct sign for the physical process in the new
channel.  Although we are dealing with an apparently classical
process, as Nelson has shown, the Brownian motion problem gives
rise to a Schr\"odinger equation, and therefore contains
properties of the differential equations of the quantum theory. We
thus see the remarkable fact that one must take into account the
physical effect of the analytic continuation of processes
occurring in a non-physical,
 in this case timelike, domain, on the total observed behavior of the
 system.
\par  In the timelike case, the velocity of the particle $\Delta w^1/
\Delta w^0 \leq 1$. We shall here use the dynamical association of
coordinate increments with energy and momentum
\begin{equation}E=M{\Delta w^0\over \Delta \tau}\,\,\,\,\,\,\, p=M{\Delta w^1\over
\Delta \tau},\label{6.2.6}\end{equation}
so that
\begin{equation} \sigma^2 = \bigl({\Delta \tau \over M } \bigr)^2
(E^2-p^2), \label{6.2.7}\end{equation}
where $M$ is a parameter of dimension mass associated with the
Brownian particle.
It then follows that  $E^2 - p^2 =\bigl({M \over \Delta \tau}
\bigr)^2\sigma^2 >0$.  For the spacelike
 case, where $p/E >1$, we may consider the transformation to an
 imaginary representation
 $E\rightarrow iE'$ and $p\rightarrow ip'$, for $E', p'$
 real\footnote[1]{ This transformation is similar to the continuation
 $p\rightarrow ip'$ in nonrelativistic tunnelling, for which the
 analytic continuation appears as an instanton.}, but $E^2 - p^2 \rightarrow p'^2 -E'^2 >0$. In this
 case, we take the analytic continuation such that the magnitude of
$\sigma^2$ remains unchanged, but can be called $-\mu^2$, so that $E'^2-p'^2=\mu^2$ with $\mu$ imaginary. The spacelike contributions are therefore obtained in this mapping by $E,p\rightarrow iE,ip$ and $\sigma \rightarrow i\mu$,
assuring the formation of the d'Alembert operator when the
 timelike and spacelike fluctuations are added with equal weight (this
equality is consistent with the natural assumption, in this case,
of an equal distribution between spacelike and timelike
contributions).  The preservation of the magnitude of the interval
reflects the conservation of a mass-like property which remains,
as an intrinsic property of the particle, for both spacelike and
timelike jumps. As mentioned before, one recalls the role of
analytic continuation in quantum field theory; for the well known Wick rotation [e.g. \cite{44}],
 however,in that case, only the 0-component is analytically continued and no clear direct
physical idea or quantity is associated with it. In the RBP the
identification of the imaginary 4-momentum is dynamical in
origin. It is due to the Lorentz structure of spacetime, which
distinguishes the transitions ${\Delta {\bf x}\over \Delta t}>1$ from those with
${\Delta{\bf x} \over \Delta t}<1$. Though one may object to the association
of $\Delta x^\mu$ with a dynamical momentum (since the instantaneous
derivative ${dx^\mu \over d\tau}$ is not defined for a Brownian
process) the Brownian motion could be understood as an approximation to a
microscopic process, just as it appears in Einstein's famous work in
1905 \cite{Einstein}, where it is assumed that the Brownian motion is produced by
collisions. The effective conservation of $E^2-{\bf p}^2$ as a real quantity
in both timelike and spacelike processes suggests that it is a
physical property which preserves its meaning in both sectors. 
\par With these assumptions,  the cross-term in
hyperbolic functions cancels in the sum, which now takes the form
\begin{equation}{\cal O} = \mu^2\bigl[{\partial^2\over \partial x^2} - {\partial^2
\over \partial t^2}\bigr] \label{6.2.8}\end{equation}
Taking into account the drift term in
Eq.~(\ref{6.2.1}), one then finds the relativistic Fokker-Planck equation
\begin{equation} {\partial \rho(x,\tau)\over \partial \tau} = \bigl\{-{\partial
 \over
 \partial
x^\mu}\beta^\mu + \langle \mu^2 \rangle {\partial \over \partial x^\mu}{\partial \over
\partial x_\mu}\bigr\} \rho(x,\tau), \label{6.2.9}\end{equation}
where $\partial/\partial x^\mu$ operates on both $\beta^\mu$ and $\rho$.
\par We see that the procedure we have followed, identifying
$\sigma^2= -\mu^2$ and assuming equal weight, permits us to construct
the Lorentz invariant \break d'Alembertian operator, as required for
obtaining a relativistically covariant diffusion equation.

 \par To see this process in terms of a higher symmetry, let us define the invariant $\kappa^2 \equiv E_t^2 -p_t^2
 \geq 0$
 for the timelike case;  our requirement is then that $E_s^2 -p_s^2 =
 -\kappa^2$
for the spacelike case.
  In the framework of a larger
 group that includes $\kappa$ as part of a three vector $(E, \kappa, p)$, the
 relation for the timelike case can be considered in terms of
 the invariant of the subgroup $O(1,2)$, i.e., $E^2 -\kappa^2 -p^2$.  The
 change in sign for the spacelike case yields the invariant
 $E^2+\kappa^2 -p^2$; we designate the corresponding  symmetry (keeping
 the order of $E$ and $p$) as $O(2,1)$.  These two groups may be
 thought of as subgroups of $O(2,2)$, where there exists a
 transformation which changes the sign of the metric of the subgroups
 holding the quantity $\kappa^2$ constant. The kinematic constraints
 we have imposed correspond to setting these invariants to zero (the
 zero interval in the $2+1$ and $1+2$ spaces).
\par The constraint we have placed on the relation of the timelike and
 spacelike invariants derives from the properties of the
 distribution function and the requirement of obtaining the d'Alembert
 operator, i.e, Lorentz covariance of the diffusion equation. It
 appears that in order for the Brownian motion to result in a
 covariant diffusion equation, the distribution function has a higher
 symmetry reflecting the necessary constraints.  The transformations
$E\rightarrow iE'$ and $p\rightarrow ip'$ used above would then
 correspond to analytic continuations from one (subgroup) sector to
 another.   We shall see a similar structure in the $3+1$ case, where
 the groups involved can be identified with the symmetries of the
 $U(1)$ gauge fields associated with the quantum Stueckelberg-Schr\"odinger
equation.
\bigskip
\bigskip
\section{The Invariance Properties of the Process}
\bigskip
\bigskip
Before formulating the 3+1 dimensional Brownian process,
 let us investigate the Lorentz invariance of the process and the
 correlation
 functions.
The averaging operations are summations with weights
(probability) assigned to each quantity in the sum. The sums in the
continuum are, of course, expressed by integrals. If we wish to assign a
relativistic Gaussian distribution function then the hyperbolic
angle integration is
 infinite unless we introduce a cutoff.
The question then arises whether our process is invariant or not.
\par We will show that we can use an arbitrary non-invariant (scalar) probability
 distribution (for example, a cutoff on the hyperbolic angle) and still
 obtain Lorentz invariant averages,
 using the imaginary representations of the `unphysical jumps'.
For example, $<\Delta w^\mu>$ stands for a summation with a scalar
weight (given by the density) over all the vectors $\Delta w^\mu$, in the domain. It is therefore a vector. Moreover under the
imaginary representation of spacelike increments relative to the
timelike ones (here we assume the timelike jumps physical), $\Delta w^\mu$ is a simple vector function over all
spacetime which has the following
 form:
\begin{eqnarray}
\Delta w^\mu&=&\Delta w'^\mu \,\, , \,\, \Delta w'^\mu
\,\,{\textrm{timelike}} \nonumber \\ 
\Delta w^\mu&=&i \Delta w'^\mu \,\, ,\,\, \Delta
w'^\mu \,\,{ \textrm{spacelike}}. \label{6.3.1}
\end{eqnarray}
where the $\Delta w'^\mu$ are real.
The quantity $<\Delta w^\mu>$ (formally written as a discrete sum) is
 given therefore by:
\begin{eqnarray}
<\Delta w^\mu>&=&\sum P(\Delta w) \Delta w^\mu=\sum_{
 \textrm{timelike}}P'(\Delta w') \Delta w'^\mu+i\sum_{\rm spacelike}P'(\Delta
 w')
 \Delta w'^\mu=\nonumber \\
  &=&<\Delta w'^\mu>_{\rm timelike}+i<\Delta
w'^\mu>_{\rm spacelike}\label{6.3.2}
\end{eqnarray}
where $P(\Delta w)$ (or $P'(\Delta w')$) is the probability(weight) of
having the vector $\Delta w$ (or $\Delta w'$). The two vectors in the
last equality in Eq.~(\ref{6.3.2}) are just normal Lorentz vectors. If we now
pick a distribution in a given frame for which the average of each of
them (independent of the other) is zero then $<\Delta w^\mu>=0$ is
true in all frames since
 the 0-vector is Lorentz invariant.

\par Building the second correlation, with the assumption of no
 correlation between spacelike jumps and timelike jumps, we find:
\begin{eqnarray}
<\Delta w^\mu \Delta w^\nu>&=&\sum P(\Delta w){\Delta
 w^\mu}{\Delta w^\nu}=\sum_{\rm timelike}P'(\Delta w') \Delta w'^\mu
 \Delta w'^\nu+\nonumber \\&+&i^2\sum_{ \textrm{spacelike}}P'(\Delta w') \Delta
 w'^\mu \Delta w'^\nu \nonumber\\&=&\sigma^{\mu \nu}_{ \textrm{timelike}}-\sigma^{\mu
 \nu}_{ \textrm{spacelike}}\propto
 \eta^{\mu \nu}D \Delta \tau  \label{6.3.3}
 \end{eqnarray}
where $\sigma^{\mu\nu}$ is the correlation tensor in each case. 
From the definition of $\Delta w'^\mu$ (a four vector) it follows that
$\sigma^{\mu \nu}$ are real Lorentz tensors.
The last equality in Eq.~(\ref{6.3.3}) is a demand that could be achieved for
the general 1+n case, by assuming that in a given frame there is an
invariant Gaussian distribution where the distribution is uniform in
all angles and that there is a cutoff in the hyperbolic angle. The
sum of the two covariant tensors (each a result of
summation on different sectors) is a Lorentz invariant tensor. The higher correlation functions do not
interest us since they are of higher order in $\rho$ and therefore in
$\Delta \tau$ and do not contribute to the Fokker-Plank
 equation.
 \bigskip
\section{The Notion of `Jumps' Versus a Continuous Process} 
\bigskip
The mapping given in Eq.~(\ref{6.3.1}) leads necessarily to a deviation from the standard mathematical formulation of Brownian motion. There the probability that a particle starting at $x$ at time $\tau$ ending at $x'$ at time $\tau'$ is equal to the probability that the particle starts at $x$ at time $\tau$ passing through any possible intermediate point $x''$ at time $\tau''<\tau'$ and going from there to the point $x'$ at time $\tau'$ . This property is expressed in the Chapman-Kolmagorov equation [e.g. \cite{26}],
\begin{equation}p(x,\tau,x',\Delta \tau')=\int_{\cal R}p(x,\tau,x'',\tau'')p(x'', \tau'',x',\tau')d^4x'', \label{6.5.1}\end{equation}      
In the relativistic formulation the vector $\Delta w'=x-x'$ could be a timelike vector therefore resulting in a real valued vector $\Delta w$ according to the mapping in Eq.~(\ref{6.3.1}) However, the two intermediate vectors $\Delta w'_1=x-x''$ and $\Delta w'_2=x''-x$ could be spacelike, and take the event out of the real manifold into a complex valued coordinate. In this case the Chapman-Kolmagorov equation does not hold, and the event may be found outside of the real manifold.
In order to build a consistent process one must adopt the concept of 'Brownian jumps' which could be a result for example of a process in which the event (similar to Einstein's original construction) undergoes collisions and for each collision, or `jump' the mapping in Eq.~(\ref{6.3.1}) holds. Therefore at each point in the physical manifold the event may take any increment spacelike or timelike (with a possible complex valued contribution to the averages). However, although the vector leading from the initial point , say $O$, to the end point, $A$, may be spacelike and therefore be represented as an imaginary vector it is understood that the event arrives at the real spacetime point $A$, never leaving the real spacetime. This structure separates the two manifolds, spacetime which is real and represents the physical coordinates of the event and a complex space representing the processes the event undergoes going from one point to another. This structure differs in that sense from the mathematical formulation due to Wiener and others, but still it can be shown that the process is invariant on the average under decomposition into shorter time subprocesses . In other words, we consider the event starting at some arbitrary point, and going for some time $\Delta \tau$. We next decompose the time interval into $M$ intervals, so that:
$$\sum_{i=1}^M \Delta \tau_i=\Delta \tau$$
We consider then the expression appearing in the Fokker-Plank equation \\
$<\Delta x^\mu \Delta x^\nu>$:
\begin{equation}<\Delta x^\mu \Delta x^\nu>=<\bigl(\sum_{i=1}^{M}\Delta x_i^\mu\bigr)\bigl(\sum_{j=1}^{M}\Delta x_i^\nu\bigr)>
\end{equation}
Since we assume that any two non-equal time jumps are not correlated, i.e. $<\Delta x_i\Delta x_j=0>$ for $i\neq j$, which leaves only the equal time averages in the sum,
\begin{equation}<\Delta x^\mu \Delta x^\nu>=\sum_{i=1}^{M}<\Delta x_i^\mu\Delta x_i^\nu>=\sigma^2\eta^{\mu \nu}\sum_{i=1}^M \Delta \tau_i=\sigma^2 \eta^{\mu \nu}\Delta \tau
\end{equation}     
where $\sigma^2$ is the diffusion constant and we used Eq.~(\ref{6.3.3}) going from the second piece in the equality to the third\par The notion of `jumps' stimulates the consideration of discrete processes, which can also be formulated within the relativistic framework and leads, under certain assumptions, to a covariant Fokker-Plank equation.
 For example let us assume a physical process in which the `jumps' occur in a very ordered way every $\tau_J$ seconds with a very small time spread (i.e. a very small probability that a collision occurs within a time different significantly from $\tau_J$).
Then, averaging the 'jumps' over a period $\tau>>\tau_J$ leads to:
\begin{eqnarray}
<\Delta w^\mu \Delta w^\nu>\cong N\sigma^2\eta^{\mu \nu}\tau_J \nonumber\\
N \tau_J<\tau<(N+1)\tau_J \label{discrete}
\end{eqnarray}
This result is due to the fact that under our assumptions during the time $\tau$, $N$ single `jumps' within separation of each other of $\tau_J$ occurred. 
The average in Eq.~(\ref{discrete})does not change when $\tau$ changes in less then $\tau_J$; however if $\tau_J$ is small then one can replace Eq.~(\ref{discrete}), with,
\begin{eqnarray}
<\Delta w^\mu \Delta w^\nu>\cong \sigma^2\eta^{\mu \nu}\tau \nonumber\\
\label{discrete1}
\end{eqnarray}  
Therefore we recover the standard result for Brownian motion. However there is one very important difference which is the fact that $\tau$ can be taken to be finitely small where in the standard Brownian process $\tau$ can be actually taken to zero.This implies that higher order derivative terms enter into the resulting 'diffusion' equation. For example for an isotropic homogeneous Gaussian distribution there will be additional even order derivative operators beyond the second order (d'Alembert) with coefficients $\sigma^n {\tau_J}^(2n-1)$ where $n$ is the (even) order of the differential operator. Since both $\tau_J$ and $\sigma^2$ are small these operators could be neglected in general, though there might be special configurations in which their effect may be significant.    
In the following we assume that the $\tau_J$ is very small compared with the macroscopic scale and that the `jumps' are practically ordered with zero spread, thus the approximation in Eq.~({\ref{discrete1}) is valid and no higher order terms are considered.
\bigskip
\section{Brownian motion in  $3+1$ dimensions}
\bigskip
\bigskip
\par In the $3+1$ case, we again separate the jumps into timelike and
 spacelike types. The spacelike jumps may be parameterized, in a given
 frame,  by
\begin{eqnarray}
\Delta w^0 &=& \mu \sinh{\alpha} \nonumber \\
 \Delta w^1 &=& \mu
 \cosh{\alpha}\cos{\phi}\sin{\vartheta} \nonumber \\
  \Delta w^2 &=& \mu\cosh{\alpha}\sin{\phi}\sin{\vartheta} \nonumber \\
   \Delta w^3 &=& \mu \cosh{\alpha}\cos{\vartheta} 
 \label{6.4.1}
 \end{eqnarray}
\par We assume  the four variables $\mu, \alpha, \vartheta, \phi$ are
independent random variables. In addition we demand in this frame that
 $\vartheta$ and $\phi$ are uniformly distributed in their ranges
 $(0,\pi)$ and $(0, 2\pi)$, respectively. In this case, we may average
 over the
 trigonometric angles, i.e., $\vartheta$ and $\phi$ and find that:
\begin{eqnarray}
<{\Delta w^1}^2>_{\phi , \vartheta}&=& <{\Delta
 w^2}^2>_{\phi,\vartheta}= <{\Delta w^3}^2>_{\phi , \vartheta}={\mu^2
 \over 3 }{\cosh}^2{\alpha}\nonumber \\
   <{\Delta w^0}^2>_{\phi ,\vartheta}&=&\mu^2 {\sinh}^2 {\alpha} \label{6.4.2}
\end{eqnarray}
We may obtain  the averages over the trigonometric angles of the
timelike jumps by replacing everywhere in Eq.~(\ref{6.4.2})
\begin{eqnarray} 
 \cosh^2{\alpha} &\leftrightarrow& \sinh^2{\alpha}\nonumber \\
  \,\,\,\,
 \mu^2 &\rightarrow& \sigma^2 
 \nonumber \end{eqnarray}
to obtain
\begin{eqnarray}
<{\Delta w^1}^2>_{\phi , \vartheta}&=& <{\Delta w^2}^2>_{\phi,\vartheta}= <{\Delta w^3}^2>_{\phi ,\vartheta}={\sigma^2 \over 3 }{\sinh}^2{\alpha}\nonumber \\
 <{\Delta w^0}^2>_{\phi , \vartheta}&=& \sigma^2{\cosh}^2
 {\alpha}, \label{6.4.3}
 \end{eqnarray}
where $\sigma$ is a real random variable, the invariant timelike interval.
Assuming, as in the $1+1$ case, that the likelihood of the jumps being
in either the spacelike or (virtual) timelike phases are equal, and
 making an
analytic continuation for which $\sigma^2 \rightarrow -\lambda^2$,
  the total average of the operator ${\cal O}$, including the
 contributions of
the remaining degrees of freedom $\mu,\lambda$ and $\alpha$  is
\begin{eqnarray}
<\cal O>&=&\bigl(<\mu^2>
<{\sinh}^2{\alpha}>-<\lambda^2><{\cosh}^2{\alpha}>\bigr){\partial^2\over
\partial t^2}+\nonumber \\
 &&{1\over 3} \bigl(<\mu^2><\cosh^2{\alpha}>-<\lambda^2><\sinh^2{\alpha}>\bigr){\bigtriangleup}
 \label{6.4.4}
 \end{eqnarray}
If we now insist that the operator $<{\cal O}>$ is invariant under
Lorentz transformations (i.e. the d'Alembertian) we impose the condition
\begin{eqnarray}
<\mu^2><{\sinh}^2{\alpha}>&-&<\lambda^2><{\cosh}^2{\alpha}>=\nonumber \\-{1\over 3} \bigl(<\mu^2><\cosh^2{\alpha}>&-&<\lambda^2><\sinh^2{\alpha}>
\bigr) \label{6.4.5}
\end{eqnarray}
Using the fact that $<\cosh^2{\alpha}>-<\sinh^2{\alpha}>=1$, and
defining \break $ \gamma \equiv <\sinh^2{\alpha}>$, we find that
\begin{equation}<\lambda^2>={1+4\gamma \over 3+4\gamma}<\mu^2> \label{6.4.6}\end{equation}
The Fokker-Planck equation then takes on the same form as in the $1+1$
case, i.e., the form Eq.~(\ref{6.2.9}).
We remark that for the $1+1$ case, one finds in the corresponding
 expression that the
$3$ in the denominator is replaced by unity, and the coefficients $4$
are
 replaced by
$2$; in this case the requirement reduces to $<\mu^2> =<\lambda^2>$
 and there is no $\gamma$ dependence.

\par We see that in the limit of a uniform distribution in $\alpha$,
for which $\gamma \rightarrow \infty,$
\begin{equation}<\lambda^2> \rightarrow<\mu^2>.\nonumber 
\end{equation}                                     
In this case, the relativistic generalization of 
nonrelativistic Gaussian distribution of the form
 $e^{-{{\bf r}^2\over dt}}$
is $e^{-{\mu^2\over d\tau}}$, which is Lorentz invariant.
\par The limiting case $\gamma \rightarrow 0$ corresponds to a
stochastic process in which in the spacelike case there are no
fluctuations in time, i.e., the process is that of a nonrelativistic
Brownian motion.  For the timelike case (recall that we have assumed
the same distribution function over the hyperbolic variable) this
limit implies that the fluctuations are entirely in the time
direction.  The limit $\gamma \rightarrow \infty$ is Lorentz
invariant,
 but
the limit $\gamma \rightarrow 0$ can clearly be true only in a
particular frame.
\bigskip
\section{ The Markov Relation and the 4D Gaussian Process}
 \smallskip
 \smallskip
 \par In developing the previous ideas leading to the formulation of a RBP, we
 assumed that the probability distribution is consistent with the Markov property
expressed in the Chapman-Kolmagorov equation [e.g. \cite{26}]
However, for the relativistic Gaussian it is not clear whether Eq.~(\ref{6.5.1}) holds. Therefore we now consider an alternative process, using the ideas developed above, resulting eventually in
the Klein-Gordon equation. Let us consider a 2D Gaussian process generated by a
distribution of the form :
\begin{equation}p(w,d\tau)={1 \over  2 \pi D d\tau } exp({-{\Delta w_0}^2-{\Delta w_1}^2 \over 2 D d\tau})
\label{6.5.2}\end{equation}
This distribution corresponds to a Markov process, a standard
normalized Wiener process, where $D$ is the diffusion constant. We now use the
coordinate representation given in Eq.~(\ref{6.2.3}) and Eq.~(\ref{6.2.3'}) for the timelike and
spacelike sectors to transform the distribution function in Eq.~(\ref{6.5.2}) in both
sectors to (use $\mu^2$ in both cases):
\begin{equation}{1 \over  2 \pi D d \tau } exp({-\mu^2 \cosh2\alpha \over 2 D d\tau})
\label{6.5.3}\end{equation}
where timelike `jumps' are physical and the measure for both sectors is $\mu d\mu d\alpha$ 
 Then, following Section 5, using Eq.~(\ref{6.3.1}), we get for the
combination of the timelike and spacelike contributions (with the appropriate sign)
of the averages, say, ${\Delta w_0}^2$,
\begin{eqnarray}
<\Delta {w_0}^2>&=& {1 \over  2 \pi D d \tau
}\int_0^{\infty}\int_{-\infty}^{\infty}\mu^3 exp({-\mu^2 \cosh2\alpha \over 2 D
d\tau})d\mu d\alpha=\nonumber \\
&=&{1 \over \pi}D d\tau \label{6.5.4}
\end{eqnarray} 
where we integrated
over $\mu$ first, using
\begin{equation}\int_0^{\infty}\mu^n exp(-a \mu^2)={\Gamma({ n+1\over
2})\over 2 a^{(n+1)/2}} \label{6.5.5}\end{equation}
 and then integrated over $\alpha$ using
\begin{equation}I_2\equiv \int_{-\infty}^{\infty}{d\alpha \over \cosh^2 2\alpha}=1 \label{6.5.6}\end{equation}
 In a similar way one finds that (using Eq.~(\ref{6.3.1}) leading to the negative sign)
\begin{equation}<{\Delta {w_1}}^2>=-{1 \over \pi}D d\tau \label{6.5.7}\end{equation}
Since the probability distribution Eq.~(\ref{6.5.2}) is symmetric in ${\Delta w_i}$ in each
sector \break $<\Delta w_0 \Delta w_1>=0$ as well as the first moments. Therefore we
get in this particular frame a d'Alembertian. However, following the results of
Section 5 we see that it is an invariant result in all Lorentz frames (though in other frames
the distribution may not appear to be Gaussian).

Next we consider the application of the 4D form of Eq.~(\ref{6.5.2})
\begin{equation}p(w,d\tau)={1 \over  4 \pi^2 D^2 (d\tau)^2 } exp({-{\Delta w_0}^2-{\Delta w_1}^2-{\Delta w_2}^2-{\Delta w_3}^2
\over 2 D d\tau}) \label{6.5.8}\end{equation} 
with measure $\mu^3d\mu cosh^2\alpha sin\theta d\theta d\alpha d\phi$ for the spacelike sector and $\mu^3d\mu sinh^2\alpha sin\theta d\theta d\alpha d\phi$ for the timelike sector. 
\par However, now calculating $<{\Delta w_0}^2>$ for the
timelike case, after averaging over the spatial angles $\theta$ and $\phi$ we find,
using Eq.~(\ref{6.4.1}),
\begin{equation}<\Delta {w_0}^2>={1 \over \pi D^2 (d\tau)^2
}\int_0^{\infty}\int_{-\infty}^{\infty}\mu^5exp({-\mu^2 \cosh2\alpha \over 2 D
d\tau})\cosh^2\alpha \sinh^2\alpha d\mu d\alpha \label{6.5.9}\end{equation} and for the spacelike
case we get the same result since the spacelike parametrization of ${\Delta w_0}^2$
is proportional to $\sinh^2\alpha$ and the spacelike volume element is proportional
to $\cosh^2\alpha$. Therefore if we use Eq.~(\ref{6.3.1}), adding the contribution of the
two sectors one obtains a complete cancellation to zero. In order to avoid this we
extend Eq.~(\ref{6.3.1}) to the form
\begin{eqnarray}
\Delta w^\mu&=&\Delta w'^\mu \,\, , \,\, \Delta w'^\mu \,\,{\rm
timelike}\nonumber \\
 \Delta w^\mu&=&i\lambda \Delta w'^\mu \,\, ,\,\, \Delta w'^\mu \,\, {\rm
spacelike} \label{6.5.10}
\end{eqnarray}

Before completing the calculation, we discuss the inclusion of the factor
$\lambda$, in Eq.~(\ref{6.5.10}). Let us consider a classical (i.e. non-stochastic) event with
a given value $m^2 \equiv \Delta w_\mu \Delta w^\mu$, moving in a timelike
direction. It then changes its state of motion and starts moving in a spacelike
direction; according to Eq.~(\ref{6.5.10}) $m^2$ changes into $\lambda^2 m^2$.
Moreover, though the event may move according to a Gaussian distribution which makes
no distinction between timelike and spacelike motions, the outcome of this motion as
represented by the $\Delta w$, in Eq.~(\ref{6.3.1}); Eq.~(\ref{6.5.10}) does distinguish the two
phases of motion. We shall see that a specific value of $\lambda$ is required for the realization of the Fokker-Planck equation.
\par The $w$ manifold is complex and it is a function of the motion on the real
manifold $w'$. Our macroscopic (physical) equations are written on the real plane of
the $w$ manifold. One can then visualize the flow of an event in spacetime similar
to a motion of a particle in a cloud chamber. There as the particle moves the
gas condenses, therefore the particle leaves a track. The track itself is not
the particle but a result of the actual motion of the particle and its interaction
with the gas in the cloud chamber.  The track in the cloud chamber is analogous to the complex representation we use for the `jumps'.  

We calculate first the expectation $<{\Delta w_0}^2>$ which is the total expectation,
summed over the timelike and spacelike sectors. Averaging over the spherical angles
$\theta,\varphi$ we get using Eq.~(\ref{6.5.9}) and Eq.~(\ref{6.5.10}),
\begin{eqnarray}
<\Delta {w_0}^2>&=&{1 -\lambda^2\over \pi} D^2 (d\tau)^2
\int_0^{\infty}\int_{-\infty}^{\infty}\mu^5exp({-\mu^2 \cosh2\alpha \over 2 D
d\tau})\cosh^2\alpha \sinh^2\alpha d\mu d\alpha =\nonumber \\
& =& {8 (1-\lambda^2)\over \pi}D
d\tau \int_{-\infty}^{\infty}{\cosh^2\alpha \sinh^2\alpha \over \cosh^3
2\alpha}d\alpha \label{6.5.11}
\end{eqnarray}
where Eq.~(\ref{6.5.5}) was used in the $\mu$ integration leading to the last equality
in Eq.~(\ref{6.5.11}).
 Using

\begin{eqnarray} \cosh^2\alpha&=& {1\over 2}(\cosh 2\alpha+1)\nonumber \\
\sinh^2 \alpha&=&{1\over2}(\cosh2\alpha-1)\label{6.5.12}
\end{eqnarray}

in Eq.~(\ref{6.5.11}) and integrating over $\alpha$ we get

\begin{equation}<\Delta {w_0}^2>={(1-\lambda^2)\over \pi}D d\tau {\pi \over 2} \label{6.5.13}\end{equation}
where we used
\begin{eqnarray}I_1&\equiv &\int_{-\infty}^{\infty}{d\alpha \over \cosh 2\alpha}={\pi \over 2} \nonumber \\
 I_3&\equiv &\int_{-\infty}^{\infty}{d\alpha \over \cosh^3 2\alpha}={\pi \over 4}
\label{6.5.14}
\end{eqnarray}

We now calculate the expectation of $<{\Delta w_1}^2>$. Averaging over the spherical
angles $\theta,\varphi$ we get, using Eq.~(\ref{6.5.9}) and Eq.~(\ref{6.5.10}),
\begin{eqnarray}
<\Delta {w_1}^2>&=&{1\over 3 \pi} D^2 (d\tau)^2
\int_0^{\infty}\int_{-\infty}^{\infty}\mu^5exp({-\mu^2 \cosh2\alpha \over 2 D
d\tau})(\sinh^4 \alpha-\lambda^2\cosh^4\alpha)d\mu d\alpha =\nonumber \\
& =& {8 \over 3 \pi}D
d\tau \int_{-\infty}^{\infty}{(\sinh^4\alpha-\lambda^2\cosh^4\alpha) \over \cosh^3
2\alpha}d\alpha \label{6.5.15}
\end{eqnarray}

Using Eq.~(\ref{6.5.12}), Eq.~(\ref{6.5.6}), Eq.~(\ref{6.5.14}), We get after integration over $\alpha$,
\begin{equation}<\Delta {w_1}^2>={D d\tau \over \pi}{1 \over 3}[(1-\lambda^2){3\pi \over
2}-4(1+\lambda^2)]\nonumber 
\end{equation}                                   
In order to obtain the d'Alembertian we insist that $<\Delta {w_1}^2>=-<\Delta
{w_0}^2>$, which leads to
\begin{equation} \lambda^2={3\pi-4 \over 3\pi+4} \label{6.5.16}\end{equation}
Finally, substituting (for example) Eq.~(\ref{6.5.16} in Eq.~(\ref{6.5.13}) We find 
\begin{equation}<\Delta w^\mu \Delta w^\nu>=\eta^{\mu \nu}{4D \over 3\pi +4}d\tau= \eta^{\mu \nu} \breve{D}d\tau \label{6.5.17}\end{equation}
where $\breve{D}$, is the actual effective diffusion constant defined by
\begin{equation} \breve{D}\equiv {4D \over 3 \pi +4} \label{6.5.18}\end{equation}
\section{Discussion and Conclusions}
\par We have constructed a relativistic generalization of Brownian               
motion, using the invariant world-time, $\tau$, to order the Brownian                     fluctuations, and separated consideration of spacelike and timelike            
 jumps to avoid the problems of negative second moments which might             
 otherwise follow from the Minkowski signature. Associating the                 
 Brownian fluctuations with an underlying dynamical process, one may think of $\gamma$ discussed in the $3+1$ case as an order parameter, where the distribution function (over $\alpha$), associated with the velocities, is determined by the temperature of the underlying dynamical system (the result for the $1+1$ case is independent of the distribution on the hyperbolic variable).  More generally it is suggestive to consider the possible thermodynamical effects of the `medium' generating the relativistic Brownian fluctuations, following similar steps taken by Einstein \cite{Einstein} in his famous work and verify whether any physical effect can be predicted.
\par  At equilibrium, where $\partial \rho/\partial \tau =0,$ the                
 resulting diffusion equation turns into a classical wave equation              
 which, in the absence of a drift term $K^\mu$, is the wave equation            
 for a massless field.  An exponentially decreasing distribution in             
 $\tau$ of the form $\exp{-\kappa \tau}$ would correspond to a                  
 Klein-Gordon equation for a particle in a tachyonic state (mass                
 squared $-\kappa$), for physical spacelike motion and for physical timelike motion to a particle with mass squared $\kappa$.                                                            
\par Choosing a cutoff in the hyperbolic angle,one finds a covariant moment and therefore,             
 covariant differential operators. However the underlying process is            
 not invariant, thus one can think of a special frame in which the              
 hyperangular distribution is uniformly distributed around 0. Boosting                
 breaks the symmetry of the hyperangular distribution, but since the the averages are tensor quantities the invariance properties are conserved, and therefore the Fokker-Plank equation (leading to the quantum equation) is invariant. This property is also used to construct the 4D Gaussian process.
 \par It was shown that a (Euclidian) Gaussian process with an appropriate 
 (weighted) complex representation for the timelike and spacelike random motions can be used to achieve the covariant quantum equation, with the assurance that it is Markovian, (it is a relativistic generalization of the Wiener process). This leads to a `cloud chamber-like' picture in which the event as it evolves leaves a track (carries a real or an imaginary phase),which is a representation of the actual motion, distinguishing the timelike and spacelike motion.
\par In the classical Stueckelberg theory the timelike (forward or backward) propagation is associated with the standard particle or antiparticle interpretation, where spacelike propagation is needed whenever one discuss classical pair creation or annihilation (with continuous passage from forward to backward motion in time). This suggests that the spacelike process may be associated with the annihilation and creation of pairs.
Moreover, though the resulting  macroscopic equation(i.e. on the level of the Fokker-Planck equation) is local and causal in the spacetime variables, the underlying microscopic process(i.e., on the level of the Brownian fluctuations) is not. It is however local and causal in $\tau$ even at the microscopic level. This non-locality in $t$ `microscopically' may lead to a mechanism providing correlations for the entangled state system.
 \par  Nelson has shown that non-relativistic Brownian motion can be associated with a Schr\"odinger equation. Equipped with the procedures we presented here, constructing relativistic Brownian motion, Nelson's methods can be generalized. One then can construct relativistic equations of Schr\"odinger (Schr\"odinger-Stueckelberg) type. The eigenvalue equations for these relativistic forms are also Klein-Gordon type equations. Moreover one can also generalize the case where the fluctuations are not correlated in different directions into the case where correlations exist, as discussed by Nelson for three dimensional Riemannian spaces. In this case the resulting equation will be a quantum equation in a curved Riemannian spacetime; as we have pointed out, the eikonal approximation to the solutions of such an equation contains the geodesic motion of     classical general relativity. The medium supporting the Brownian motion may be identified with an ``ether''  for which the problem of local Lorentz symmetry is solved. 
 This study opens up several tracks of possible research. Nelson, discussing the E.P.R  \cite{Nelson1985} system confronted the fact that such a system may be either described by a non-local Markov process or a local non-Markov process. The Markov process is simple to implement but Nelson was disturbed by the introduction of non-local interaction. However, the non-Markov process is very difficult to apply. The RPB developed here may bridge the two possibilities since an ordered (causal) Markov process in $\tau$ may appear to be a non-Markovian (or possibly non-local and certainly non causal) process in $t$. For example for the Gaussian process, looking for the probability of finding the event changing its spatial position $\Delta x$ after $\Delta t$ has passed, one may integrate Eq.~(\ref{6.5.2}) over all $\tau$. This results however, in ${{1 \over {\Delta x}^2+{\Delta t}^2}}$ which is not integrable and therefore can not be normalized. This however is not surprising, since the probability of finding the event in $\Delta x$ after $\Delta t$ is not well defined (there may be several values of $\Delta x$ for a given $\Delta t$). For example the number of particles, as in Stueckelberg original construction \cite{Stueckelberg} depends on the trajectory through which the point is reached. Defining an appropriate one particle probability resulting from the initial process occurring in $\tau$ demands a restriction of the sample space before integrating over $\tau$ i.e., using the conditional probability restricted to processes for which the event's $t$ coordinate is monotonic in $\tau$ (no pairs are created).     
\bigskip 
\par Finally we would like to point out that generating a covariant quantum equation through an RBP leads to a possible relation between quantum mechanics and gravitation. In the context of this work, the metric of gravity can appear as an anisotropy  in the correlations  that lead to quantum equations for which the ray, or eikonal approximation, corresponds to the classical geodesic flow of general relativity. It furthermore appears interesting to generalize Einstein's famous work on this process introducing thermodynamic concepts to the resulting geometrical structure of the theory.
\bigskip
\section{Acknowledgement}
\bigskip
\par One of us (L.P.H.) would like to thank the Institute for Advanced Study, Princeton, N.J. for partial support, and Steve Adler for his hospitality, during his visit in the Spring Semester (2003) when much of this work was done. He also wishes to thank Philip Pearle for helpful discussions.

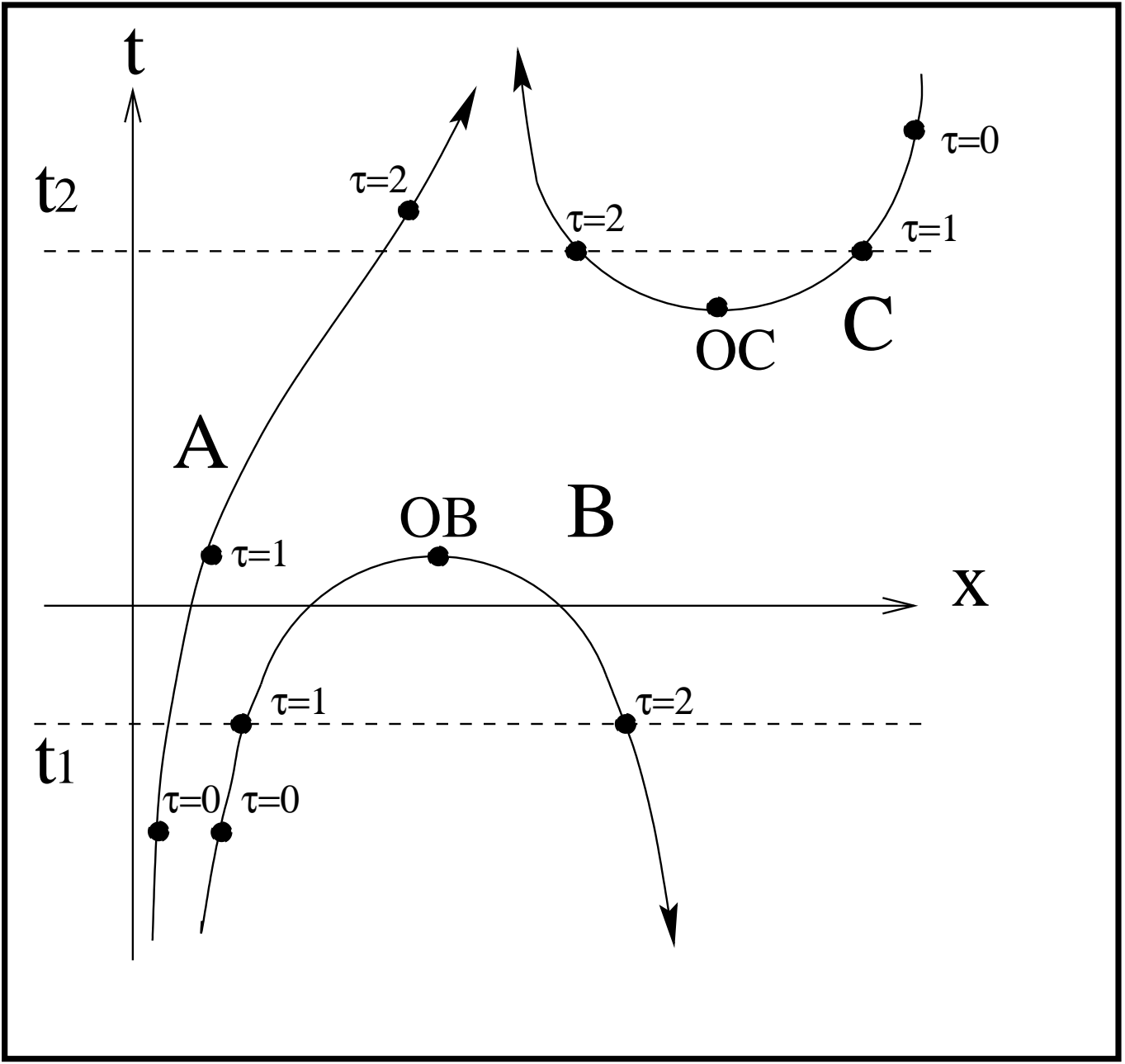
\end{document}